\documentclass[10pt, twocolumn, pre, aps, superscriptaddress, showpacs]{revtex4-1}
\usepackage{amsmath, graphicx, subfigure, tikz}
\begin{document}

\title{Reentrant Ferromagnetic Ordering of the Random-Field Heisenberg Model\\
in $d>2$ Dimensions: Fourier-Legendre Renormalization-Group Theory}

\author{Alpar T\"urko\u{g}lu}
    \affiliation{Department of Physics, Bo\u{g}azi\c{c}i University, Bebek, Istanbul 34342, Turkey}
    \affiliation{Department of Electrical and Electronics Engineering, Bo\u{g}azi\c{c}i University, Bebek, Istanbul 34342, Turkey}
\author{A. Nihat Berker}
    \affiliation{Faculty of Engineering and Natural Sciences, Kadir Has University, Cibali, Istanbul 34083, Turkey}
    \affiliation{T\"UBITAK Research Institute for Basic Sciences, Gebze, Kocaeli 41470, Turkey}
    \affiliation{Department of Physics, Massachusetts Institute of Technology, Cambridge, Massachusetts 02139, USA}

\begin{abstract}
The random-magnetic-field classical Heisenberg model is solved in spatial dimensions $d\geq 2$ using the recently developed Fourier-Legendre renormalization-group theory for $4\pi$ steradians continuously orientable spins, with renormalization-group flows of 12,500 variables. The random-magnetic-field Heisenberg model is exactly solved in 10 hierarchical models, for $d=2,2.26,2.46,2.58,2.63,2.77,2.89,3$. For non-zero random fields, ferromagnetic order is seen for $d > 2$.  This ordering shows, at $d = 3$, reentrance as a function of temperature.
\end{abstract}
\maketitle

\section{Heisenberg Spins, Lower-Critical Dimension, Random Magnetic Fields}

Random magnetic fields and Heisenberg spins (n=3 components, $4\pi$ steradians continuously orientable) constitute a double challenge to ordering under quenched randomness and varying spatial dimensions. In ordering under quenched randomness, in the previous problem of random-magnetic-field $n=1$ component Ising spins ($\pm 1$ discretely orientable), after an intense experimental and theoretical controversy between lower-critical spatial dimension $d_c=2$ claims \cite{Jaccarino,Wong,BerkerRandH} and $d_c=3$ claims \cite{Birgeneau}, the issue was settled for $d_c=2$.\cite{Machta,Falicov}  That $d_c$ is not 3 fell in contradiction with the prediction of a dimensional shift of 2 due to random fields, coming from all-order field-theoretic expansions from $d=6$ down to $d=1$ \cite{Aharony}, which indeed is a considerable distance to expand upon for a small-parameter expansion of $\epsilon=6-d$.

In ordering under varying spatial dimensions $d$, direct position-space renormalizaton-group theory has been successful across the board in determining the lower-critical dimension $d_c$, below which no ordering occurs, for all uniform systems and complex quenched random systems.  These renormalization-group studies have indeed yielded $d_c=1$ for the $n=1$ component Ising model \cite{Migdal,Kadanoff}, $d_c=2$ for the $n=2$ XY model \cite{Jose} (this study also yielding the low-temperature critical phase at $d=2$), and $2<d_c<3$ for the $n=3$ Heisenberg model \cite{Tunca1}.  Including the complexity of quenched randomness, these studies have yielded $d_c=2$, as mentioned above, for the random-field Ising model \cite{Machta,Falicov}, $3.81<d_c<4$ for the random-field XY model \cite{Kutay} with a critical line at zero temperature, in fact a non-integer $d_c=2.46$ for the Ising spin-glass with random ferromagnetic and antiferromagnetic bonds \cite{Atalay}, and $2<d_c<3$ for the Heisenberg spin-glass \cite{Tunca2}, the latter actually revealing a nematic phase, namely the occurrence of a liquid-crystal phase in a dirty magnet. These renormalization-group calculations have also, for example, shown chaos inherent in spin-glass phases \cite{McKayChaos,McKayChaos2,BerkerMcKay}, the finite-temperature phase diagram of high-$T_c$ superconductors \cite{highTc}, the changeover from first-order to second-order phase transitions under random bonds \cite{HuiBerker,erratum}, and the occurrence of first- and second-order phase transitions as a function of number of states $q$ in Potts models \cite{Nienhuis,Andelman,Devre}.

In this study, a logically next step is taken, in studying the random-magnetic-field Heisenberg spins, with n=3 components, continuously orientable in $4\pi$ steradians, using the recently developed Fourier-Legendre renormalization-group theory \cite{Tunca1,Tunca2}. The random-magnetic-field Heisenberg model is exactly solved in 10 hierarchical models, for $d=2,2.26,2.46,2.58,2.63,2.77,2.89,3$. Under non-zero random fields, ferromagnetic order is seen for $d > 2$.  This ordering shows, at $d = 3$, disorder-order-disorder phase reentrance as a function of temperature.
\begin{figure}[ht!]
\centering
\includegraphics[scale=0.5]{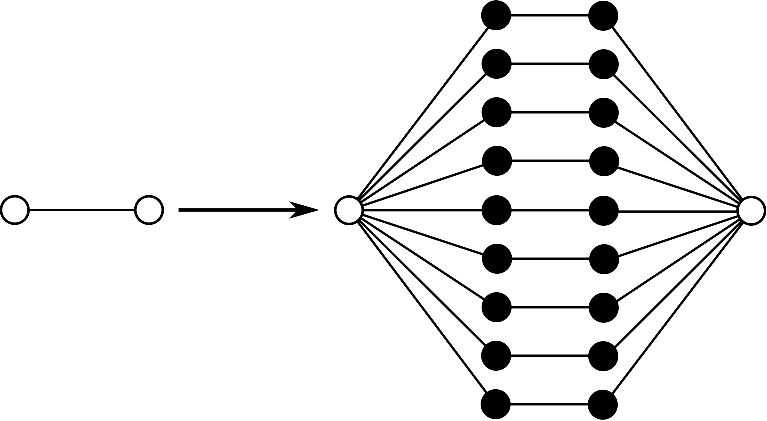}
\caption{From Ref.\cite{Artun}: Construction of a $d=3$ hierarchical model used in this study.  A hierarchical model is constructed by repeatedly self-imbedding a graph into each of its bonds.  The random-magnetic-field Heisenberg model is exactly solved in 10 hierarchical models in this study, for $d=2,2.58,3$ with $b=2$ and $d=2,2.26,2.46,2.63,2.77,2.89,3$ with $b=3$, where $b$ is the length rescaling factor, namely the number of bonds between the external (open circle) sites. The exact solution of a hierarchical model proceeds in the opposite direction of its construction \cite{BerkerOstlund,Kaufman1,Kaufman2,BerkerMcKay}.}
\end{figure}
\begin{figure}[ht!]
\centering
\includegraphics[scale=0.2]{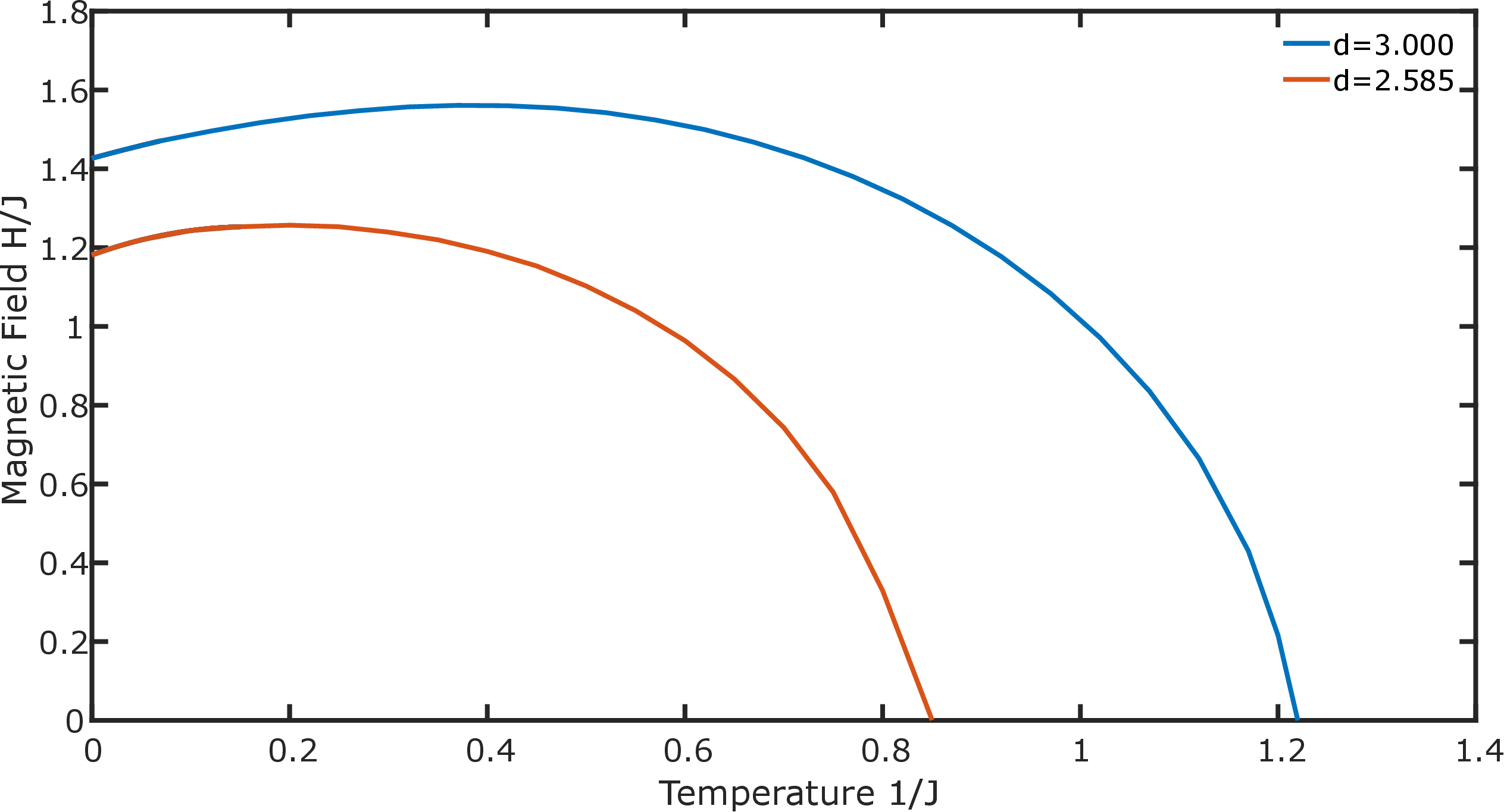}
\caption{The calculated phase diagrams of the random-field Heisenberg model, for $d=2.28$ and $d=3$ (outer curve).  The exact solutions of a $b=2$ hierarchical lattice yield these results.  The calculation shows that no ordering occurs in $d=2$.}
\end{figure}
\section{Fourier-Legendre Renormalization Group}
The random-field Heisenberg model is defined by the Hamiltonian
\begin{equation}
- \beta {\cal H} = J \sum_{\left<ij\right>} \, \vec s_i\cdot \vec s_j + \sum_{\left<ij\right>} \, \vec H_j\cdot \vec s_j,
\end{equation}
where the classical spin $\vec s_i$ is the unit spherical vector at lattice site $i$ and the sums $<ij>$ are over all
nearest-neighbor pairs of sites. In the second term, $\vec H_j$ are magnetic fields that are frozen in random directions.  In our model, the random magnetic field is attached to every site, counting from its bond coming from the left, as given in Eq. (1).  We take constant magnitude, $|\vec H_i| = H$, but random directions in $4\pi$ steradians. (This condition is of course not conserved under renormalization group.)  Note that the dimensionless $J$ and $H$ include a division by temperature, namely the factor $\beta=1/k_{B}T$.  We solve this model on the hierarchical lattice, as shown in Fig. 1.  The formulation of exactly soluble hierarchical models \cite{BerkerOstlund,Kaufman1,Kaufman2,BerkerMcKay} yielded a plethora of exactly soluble models custom-fitted to the physical problems on hand.\cite{Clark,Kotorowicz,ZhangPP,Jiang,Derevyagin2,Chio,Teplyaev,Myshlyavtsev,Derevyagin,Shrock,Monthus,Sariyer}
The hierarchical model that we use, for length-rescaling factors $b=2,3$, is the original $d=2,b=2$ hierarchical model, introduced in Fig. 2(c) of \cite{BerkerOstlund} in 1979 and is a member of the most used family of hierarchical models, namely the so-called "diamond" family.  We solve the random-field Heisenberg problem in this model, for dimensions $d=2,2.26,2.46,2.58,2.63,2.77,2.89,3$.

\begin{figure}[ht!]
\centering
\includegraphics[scale=0.2]{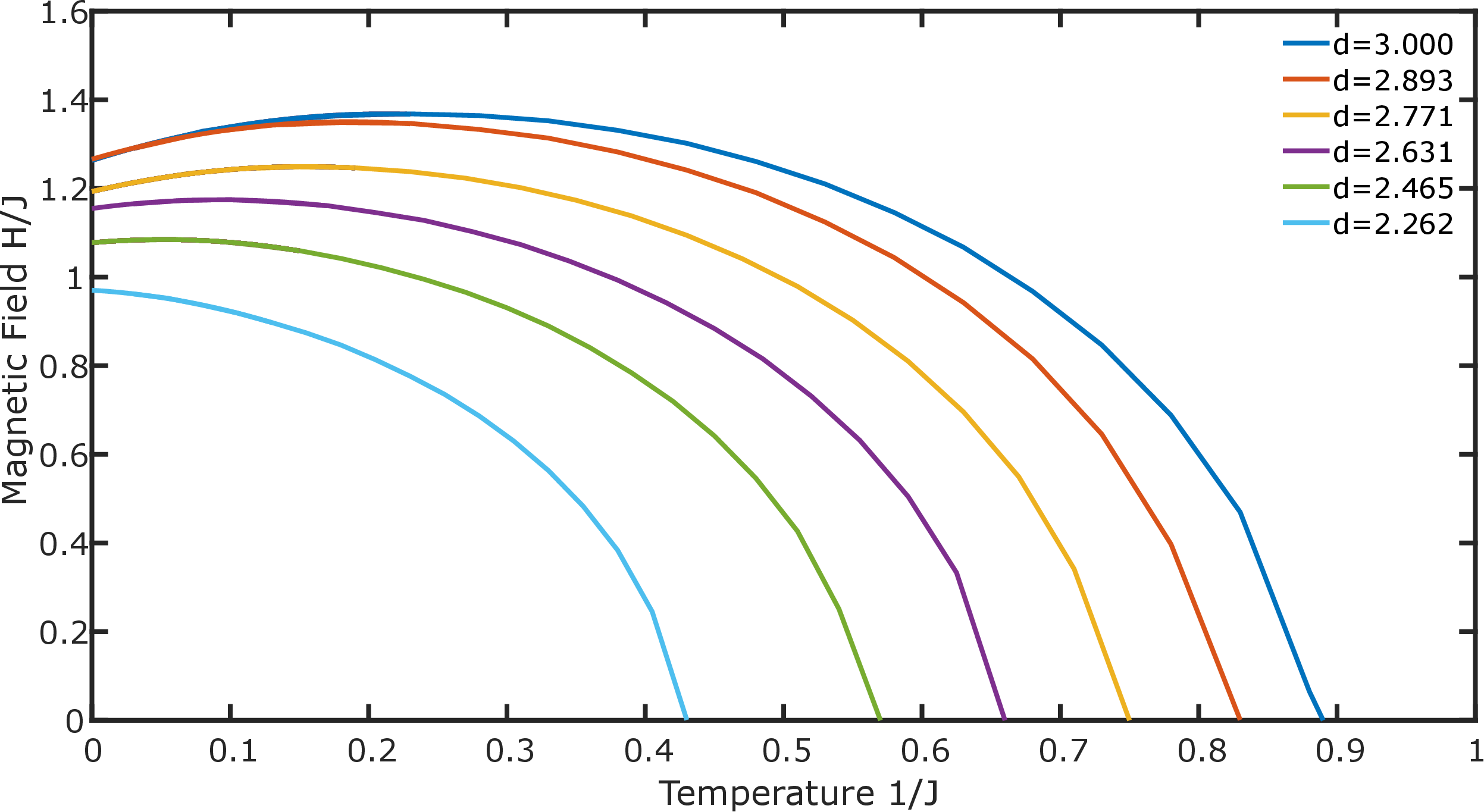}
\caption{Six calculated phase diagrams of the random-field Heisenberg model, for $d=2.26,2.46,2.63,2.77,2.89,3$ (inside to outer curves).  The exact solutions of a $b=3$ hierarchical lattice yield these results.  The calculation shows that no ordering occurs in $d=2$. Starting at $d=2.46$, temperature reentrance occurs and is magnified as $d=3$ is approached. }
\end{figure}

The solution of a hierarchical model proceeds in the opposite direction of its construction.  At each scale change, namely renormalization-group step, the spins on the internal sites (shown with black circles in Fig. 1) are eliminated by integrating, in the partition function, over their directions continuously varying over the unit sphere with angle $4\pi$ steradians, thus generating renormalized direct interaction between the spins on the outer sites (shown with open circles in Fig. 1).  This procedure involves decimation, namely the integration over the intermediate spin in two consecutive bonds in series, and the bond addition of two bonds connected in paralel to the same two sites.  The derivations for each of these two operations are given in Refs. \cite{Tunca1,Tunca2}

\begin{figure*}[ht!]
\centering
\includegraphics[scale=0.35]{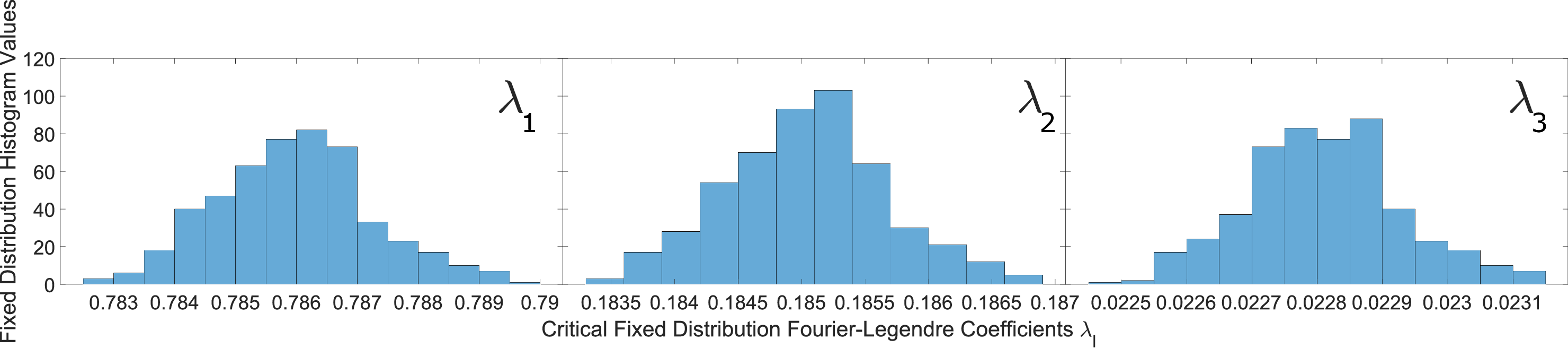}
\caption{The fixed distribution, unstable under renormalization group, controlling the phase boundary between the ferromagnetic and disordered phases in $d=3$ for $b=2$. The unstable critical fixed distributions of the Fourier-Legendre coefficients $\lambda_1,\lambda_2,\lambda_3$ are shown here.  As explained in the text, at this fixed point, $\lambda_0 = 1$.  The fixed distributions of the 21 other Fourier-Legendre coefficients $\lambda_{4-24}$ entering our calculations are not shown here.}
\end{figure*}

The exponentiated nearest-neighbor Hamiltonian between sites $(i,j)$ is expanded in terms of the Fourier-Legendre series,
\begin{equation}
 u_{ij}(\gamma) = e^{- \beta {\cal H}_{ij}(\vec s_i,\vec s_j)} = \sum_{l=0}^{\infty}{\lambda_l^{(ij)} P_l(\cos \gamma)}
\end{equation}
where $P_l(\cos \gamma)$ are the Legendre polynomials and $\gamma$ is the angle between the unit vectors $(\vec s_i,\vec s_j)$.  The expansion coefficients $\lambda_l$ are determined with
\begin{equation}
\lambda_l^{(ij)} = \frac{2l+1}{2}\int_{-1}^{1}{u_{ij}(\gamma)P_{l}(\cos \gamma)\,d(\cos \gamma)}.
\end{equation}
For decimation,
\begin{equation}
\tilde{u}_{13}(\gamma_{13}) = \int u_{12}(\gamma_{12}) u_{23}(\gamma_{23}) \frac {d\vec s_2}{4\pi},
\end{equation}
a simple equation has been derived \cite{Tunca1,Tunca2},
\begin{equation}
\tilde{\lambda}_l^{(13)} = \frac{\lambda_{l}^{(12)}\lambda_{l}^{(23)}}{2l+1} \,,
\end{equation}
where tilda denotes decimated. This procedure is repeated until the length-rescaling factor $b$ is obtained, namely until the $b$ bonds in series are replaced by one decimated bond.  For adding two bonds $A$ and $B$ between sites $(i,j)$,
\begin{equation}
\tilde{u}'_{ij}(\gamma_{12}) = \tilde{u}_{ij}^A(\gamma_{12})\tilde{u}_{ij}^B(\gamma_{12}),
\end{equation}
where prime denotes added, a Fourier-Legendre equation has also been derived \cite{Tunca1,Tunca2},
\begin{equation}
\tilde{\lambda}_{l}'=
=\sum_{l_1=0}^{\infty}\sum_{l_2=0}^{\infty}\tilde{\lambda}_{l_1}^A\tilde{\lambda}_{l_2}^B{\langle l_1 l_2 0 0 | l_1 l_2 l 0\rangle}^2,
\end{equation}
where the bracket notation is the Clebsch-Gordan coefficient with the restrictions $l_1+l_2+l=2s, s\in \mathbf{N} $; $|l_1-l_2|\leq l\leq |l_1+l_2|$.  This procedure is repeated until the $b^{d-1}$ bonds in parallel are combined, yielding the renormalized interaction $u'_{ij}$ between the outer spins (open circles) in the graph.  Thus, the renormalization-group flows are in terms of the Fourier-Legendre coefficients $\lambda_l'(\{\lambda_l\})$.  With no approximation, after every decimation and after setting up the initial conditions, the coefficients $\{\lambda_l\}$ are divided by the largest $\lambda_l$.  This is equivalent to subtracting a constant term from the Hamiltonian and prevents numerical overflow problems in flows inside the ordered phase.  We have kept up to $l=24$ in our numerical calculations of the trajectories.

The renormalization-group trajectories are effected by repeated applications of the above transformation.  The initial points of these trajectories are obtained numerically effecting Eq.(2), obtaining $u_{ij}$ for 500 different random fields.  At every step of the renormalization-group transformation, by randomly grouping $b^d$ unrenormalized $u_{ij}$, we generate one renormalized $u'_{ij}$, 500 times.  Thus, since each $u_{ij}$ is defined by 25 Fouries-Legendre coefficients, our renormalization-group flows are in the (large) space of 12,500 coefficients.

\section{Renormalization-Group Flows of the Fourier-Legendre Coefficients and Phase Transitions}

Under repeated applications of the renormalization-group transformation of Sec. II, the Fourier-Legendre coefficients (FLC) flow to a stable fixed point, which is the sink of a thermodynamic phase.  The sink of the disordered phase has $\lambda_0 = 1$ and all other FLC equal to zero, $\lambda_{l>0} = 0$, meaning a constant $u$ that is not dependent on $\gamma$, namely a non-interacting system at the sink.  This sink attracts all points of the disordered phase, which it epitomizes.  In $d=2$ for $H>0$, the disordered sink is the only sink and therefore the disordered phase is the only thermodynamic phase of the random-field system.

For $d>2$, another sink also occurs with the FLC non-zero and proportional to $2l+1$, making $u(\gamma)$ a delta function at zero angular separation of the spins as also seen in Refs.\cite{Tunca1,Tunca2}.  This is the sink of the low-temperature ferromagnetic phase.  The disordered sink continues, as the sink of the high-temperature disordered phase.  The boundary of critical points (Figs. 2 and 3) between these two phases is controlled by an unstable fixed distribtion, shown in Fig. 4.  The unstable critical fixed distributions of the Fourier-Legendre coefficients $\lambda_1,\lambda_2,\lambda_3$ are shown in Fig. 4.  At this fixed point, $\lambda_0 = 1$.  The fixed distributions of the 21 other Fourier-Legendre coefficients entering our calculations are not shown here.

\section{Phase Diagrams and Phase Reentrance}
The calculated phase diagrams, for eight different spatial dimensions $d =2,2.26,2.46,2.58,2.63,2.77,2.89,3$, are shown in Figs. 2 and 3.  The non-integer dimensions are constructed by varying, in the hierarchical model (Fig. 1), the number of parallel strands, which is equal to $b^{d-1}$.  There is no ordered phase, under random fields, for $d=2$.   The ferromagnetic phase persists, up to a temperature-dependent random-field strength, for all other studied dimensions.  The intercepts of the phase boundaries, for zero field and zero temperature, are given in Figs. 5 and 6.

\begin{figure}[ht!]
\centering
\includegraphics[scale=0.2]{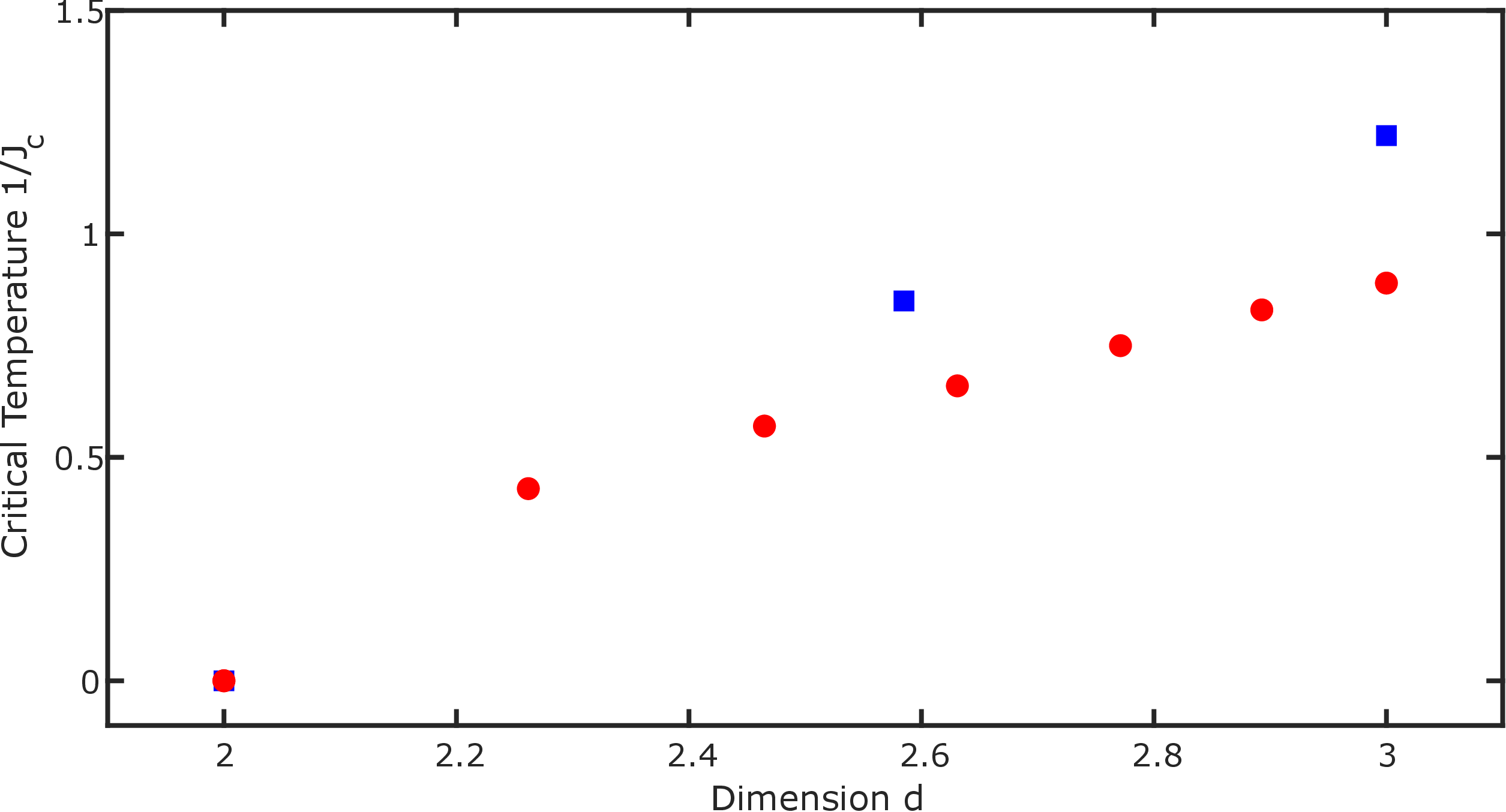}
\caption{The calculated zero-field critical temperatures with respect to spatial dimension $d$, for the $b=2$ (squares) and $b=3$ (circles) hierarchical models.}
\end{figure}
\begin{figure}[ht!]
\centering
\includegraphics[scale=0.2]{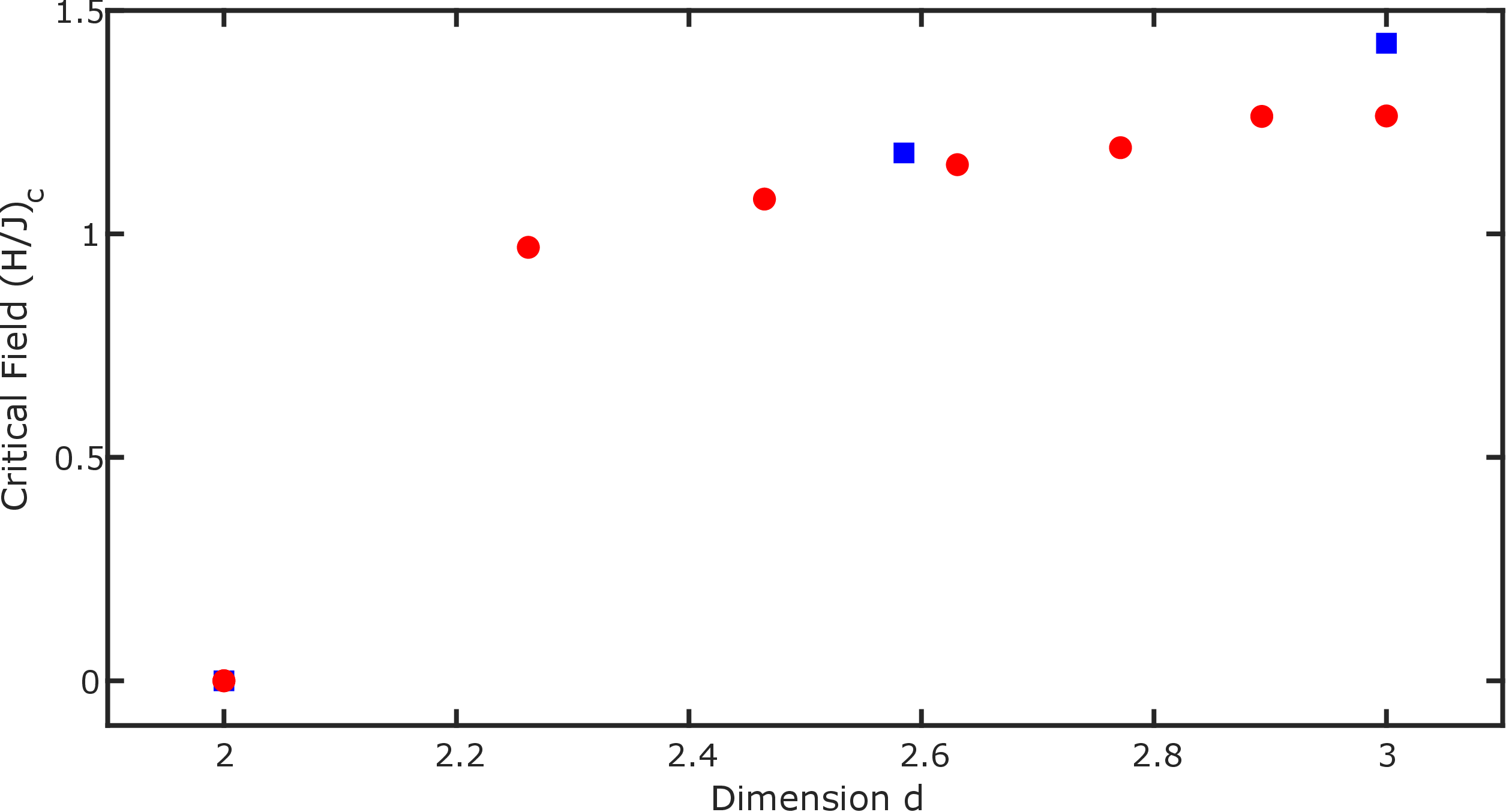}
\caption{The calculated zero-temperature critical fields with respect to spatial dimension $d$, for the $b=2$ (squares) and $b=3$ (circles) hierarchical models.}
\end{figure}

The ferromagnetic-disordered phase boundary shows temperature reentrance in $d = 2.46,2.58,2.63,2.77,2.89,3$, namely, as temperature is lowered at constant random field $J/H$ (with temperature divided out), the system goes, as usual, from the disordered phase to the ordered ferromagnetic phase. However, as the temperature is further lowered, the system goes from the ordered ferromagnetic phase back to the disordered phase. Reentrance is the reversal of a thermodynamic trend as the system proceeds along one given thermodynamic direction.  Since its observation in liquid crystals by Cladis \cite{Cladis}, this at-first-glance strange phenomenon has attracted attention by the need for a physical mechanistic explanation, which has been disparate in disparate systems. Thus, in liquid crystals the explanation has been the relief of close-packed dipolar frustration by positional fluctuations (librations) \cite{Netz,Garland}, in closed-loop binary liquid mixtures the explanation has been the
asymmetric orientational degrees of freedom of the components \cite{Walker}, in surface adsorption the explanation has been the buffer effect of the second layer \cite{Caflisch}. In spin-glasses, a magnetic system with quenched randomness as the random-field system studied here, and where there is orthogonally bidirectional reentrance, the effect of frustration in both disordering and changing the nature of ordering (to spin-glass order) is the cause \cite{Ilker1}. In cosmology, reentrance is due to high-curvature (black hole) gravity \cite{Mann1, Mann2}.  In Potts and clock model interfacial densities, in lowering the temperature, when the system orders in favor of state $a$, the preponderance of the latter also increases its interface with the other states.  However, as this preponderance further increases and in fact takes over the system, the other states are eliminated and their interface with $a$ thus is also eliminated.\cite{Artun}  In the current random-field Heisenberg model, we see that at intermediate temperatures the spins under random fields heal with the overall ferromagnetic order direction, but at lower temperatures break the system into domains dictated by local random fields, destroying long-range order.

\section{Conclusion}
We have solved, on 10 different hierarchical latttices and eigth different spatial dimensions $d =2,2.26,2.46,2.58,2.63,2.77,2.89,3$, the random-field Heisenberg model, using the newly developed Fourier-Legendre renormalization group, following the global renormalization-group trajectory of 12,500 Fourier-Legendre coefficients.  For $d =2.26,2.46,2.58,2.63,2.77,2.89,3$, the ordered ferromagnetic phase persists, up to a temperature-dependent threshold field strength.  This is shown in the calculated phase diagrams (Figs. 2 and 3). For $d = 2.46,2.58,2.63,2.77,2.89,3$, the phase diagrams show reentrance, in the sense that the phases disordered-ordered-disordered are encountered as temperature is lowered.  This calculated result has a phenomenological explanation.

\begin{acknowledgments}
Support by the Academy of Sciences of Turkey (T\"UBA) is gratefully acknowledged.  We are grateful to Egemen Tunca for very useful discussions.
\end{acknowledgments}

\end{document}